\documentclass[11pt,oneside]{article}
\usepackage{graphicx}
\begin{document}
 
\title{{\bf FORMATION OF GALACTIC SYSTEMS IN LIGHT OF THE MAGNESIUM ABUNDANCE
IN FIELD STARS: THE THICK DISK}}
\author{{\bf V.~A.~Marsakov, T.~V.~Borkova}\\
Institute of Physics, Rostov State University,\\
194, Stachki street, Rostov-on-Don, Russia, 344090\\
e-mail: marsakov@ip.rsu.ru, borkova@ip.rsu.ru}
\date{accepted \ 2005, Astronomy Letters, Vol. 31 No. 8, P.515-527 }
\maketitle

\begin {abstract}
The space velocities and Galactic orbital elements of stars 
calculated from the currently available high-accuracy observations 
in our summary catalog of spectroscopic magnesium abundances
in dwarfs and subgiants in the solar neighborhood are used to 
identify thick-disk objects. We analyze the relations between chemical, 
spatial and kinematic parameters of F--G stars in the identified 
subsystem. The relative magnesium abundances in thick-disk 
stars are shown to lie within the range $0.0 <[Mg/Fe]<0.5$ and 
to decrease with increasing metallicity starting from 
$[Fe/H]\approx -1.0$. This is interpreted as evidence
for a longer duration of the star formation process in the 
thick disk. We have found vertical gradients in metallicity 
(grad$_Z[Fe/H]= -0.13 \pm 0.04$~kpc$^{-1}$) and relative 
magnesium abundance (grad$_Z [Mg/Fe]= 0.06 \pm 0.02$~kpc$^{-1}$),
which can be present in the subsystem only in the case of its 
formation in a slowly collapsing protogalaxy. However the 
gradients in the thick disk disappear if the stars whose orbits 
lie in the Galactic plane but have high eccentricities and 
low azimuthal space velocities atypical of the thin-disk
stars are excluded from the sample. The large spread in 
relative magnesium abundance ($-0.3 <[Mg/Fe]< 0.5$) in the 
stars of the metal-poor "tail"of the thick disk which constitute 
$\approx 8\,\%$ of the subsystem, can be explained in terms 
of their formation inside isolated interstellar clouds that 
interacted weakly with the matter of a single protogalactic 
cloud. We have found a statistically significant negative 
radial gradient in relative magnesium abundance in the 
thick disk(grad$_R [Mg/Fe]= -0.03 \pm 0.01$~kpc$^{-1}$) 
instead of the expected positive gradient. The smaller 
perigalactic orbital radii and the higher eccentricities 
for magnesium-richer stars, which among other stars, are 
currently located in a small volume of the Galactic space 
near the Sun are assumed to be responsible for the gradient 
inversion. A similar but statistically less significant 
inversion is also observed in the subsystem for the 
radial metallicity gradient.
\\


\end {abstract}

\section*{Introduction}

An insufficient accuracy of determining the ages
of not only single stars, but also huge stellar ensembles,
globular and open clusters, which are believed to consist 
of coeval stars, is a serious obstacle to reconstructing 
the formation history of the structure in the early Galaxy.
Here, a "chemical" age indicator could be a help. According 
to present views, all of the chemical elements heavier 
than boron have been produced in nucleosynthesis reactions 
in stars of various masses. It thus follows that the number 
of atoms of heavy elements will inevitably increase as 
the Galaxy evolves, and the ratio of the number of atoms 
of heavy elements to the number of hydrogen atoms in the
atmospheres of unevolved stars (i.~e. their total metallicity) 
must serve as an indicator of their age. However numerous 
studies show that no age--metallicity relation can be 
traced for old Galactic populations convenient to use the 
relative abundances of various chemical elements instead 
of the total metallicity as a chemical age indicator of 
stars, since the formation of most elements can be 
attributed rather reliably to a particular nucleosynthesis 
process in stars of certain masses that evolve in a 
theoretically determined time. In particular high-mass 
type II supernovae (SNe\,II) are the main suppliers of 
$\alpha$-capture and r-process elements and a small amount 
of iron-group elements to the interstellar medium. In contrast, 
the bulk of the iron-group elements are synthesized in 
lower mass stars that are members of close binaries and that 
explode as type Ia supernovae (SNe\,Ia). The $\alpha$-elements 
are produced in a shorter time than iron, because the 
evolution times of SNe\,II ($\approx 30$~Myr) and SNe\,Ia 
($\approx 1-2$~Gyr) differ (see, e.~g. Matteucci and Greggio 
1986; Thielemann et~al.1990; Tsujimoto et~al.1995; 
Matteucci 2001)(although these authors believe that isolated,
the very first SNe\,Ia could appear considerably earlier 
than this time only $\approx 0.5$~Gyr later). Since the 
contribution of SNe\,Ia to the synthesis of iron-group elements 
is significantly larger than their contribution to the synthesis 
of $\alpha$-elements, the [$\alpha$/Fe] ratio will decrease 
in the Galaxy as the interstellar medium is enriched 
with the remnants of these supernovae. Thus at least 
$\approx 1$~Gyr after the initial burst of star formation 
will pass by the time [$\alpha$/Fe] will begin to decrease.

It is well known that the boundary between the halo and the 
thick disk can be drawn through the point $[Fe/H]\approx -1.0$.
Indeed, the distributions of subdwarfs, RR~Lyrae variable 
stars, and globular clusters in heavy-element abundance 
exhibits a clear deficit of objects in the vicinity of 
precisely this point (see, e.~g. Marsakov and Suchkov 1977;
Zinn 1985; Borkova and Marsakov 2000, 2002). At the same time, 
there is evidence that the knee in the [$\alpha$/Fe]--[Fe/H] 
relation also occurs approximately at this location in our 
Galaxy. Never the less, there is no consensus about whether 
the relative abundance of $\alpha$-elements decreases with 
increasing total metallicity inside the thick disk. In particular 
having analyzed their rather small samples of stars, Gratton
et~al.(2000) and Fuhrmann (2000) argue that the abundance of 
$\alpha$-elements in thick-disk stars is approximately the 
same and decreases abruptly as one goes to thin-disk stars. 
At the same time, Prochaska et~al.(2000) and Feldzing et~al.(2003)
derived the [$\alpha$/Fe]--[Fe/H] relation inside the thick
disk from their small samples, with the knee occurring precisely
at the point $[Fe/H ]\approx -1.0$. The difference between
these two results is of fundamental importance in correctly 
reconstructing the formation history of the thick disk. Indeed 
if the thick disk were formed during the collapse of a protogalactic 
cloud then a trend in the relative abundance of $\alpha$-elements 
with increasing heavy-element abundance would inevitably arise in
it (Prochaska et~al. 2000). The thin disk could subsequently be 
formed in the Galaxy as a result of the ongoing collapse (see 
Burkert et~al. 1992). The dificulty of this model lies in the 
fact that the total collapse time scale of the protogalactic 
cloud is much shorter than the evolution time scale of SNe\,Ia
progenitors. This dificulty can be easily circumvented in a 
different class of models where the thick disk is formed from 
the very first thin-disk stars through the interaction of the 
Galaxy with a close satellite galaxy (see Kroupa 2002, Quinn 
et~al. 1993). If further the stars of the newly forming thin disk
are assumed to be formed from interstellar matter with additions 
of accreted intergalactic metal-poor gas, then the existence of 
metal-poor stars with thin-disk kinematics and the metallicity 
overlap between thick- and thin-disk stars in the range 
$-0.8 <[Fe/H ]< -0.3$ can be explained (Fuhrmann et~al. 1995;
Fuhrmann 1998). The dificulty of this model lies in explaining 
the existence of metal-rich globular clusters in the thick
disk (Marsakov and Suchkov 1977). They cannot be heated as 
stars by a satellite galaxy and can appear only as a result 
of the star formation that accompanies this interaction 
(Gratton et~al. 2000).

Magnesium is the best-studied $\alpha$-element since it has 
visible lines of various intensities and degrees of excitation 
in unevolved F-–G stars. According to present views, almost 
all of the magnesium is synthesized in the envelopes of 
high-mass ($M>10_\odot M$) presupernovae with hydrostatic 
carbon burning and is subsequently ejected into the interstellar 
medium by SNe\,II explosions. To perform a comprehensive 
statistical analysis of the chemical, physical, and 
spatial-kinematic properties of stellar populations in
an effort to reconstruct the formation and evolution
history of the Galactic subsystems, we compiled a summary 
catalog of spectroscopic magnesium abundances in stars with 
accurate parallaxes (Borkova and Marsakov 2005). In this 
catalog, we collected almost all of the magnesium abundances 
in dwarfs and subgiants in the solar neighborhood published 
in the past fifteen years and calculated the space velocity
components and Galactic orbital elements for them.
The size of our catalog is several fold larger than the
size of any sample that has been used so far to analyze
the chemical evolution of the Galaxy; this allows us
to largely get rid of the observational selection that is
often inherent in smaller samples. In this paper we
consider the stellar population of the Galactic thick
disk.

\section*{INPUT DATA.}

In our compiled catalog the effective temperatures, surface 
gravities, and [Fe/H] values for each star were obtained 
by averaging some 2000 determinations from 80 publications.
The accuracies of the mean values of these quantities 
estimated from the scatters of the deviations of individual 
determinations about the calculated means for each star of the
sample are $\varepsilon Ò_{eff}= \pm 56^\circ$ and $\pm 82^\circ$, 
$\varepsilon$~log~$g = \pm 0.12$ and $\pm 0.24$, and 
$\varepsilon [Fe/H]= \pm 0.07$~dex and $\pm 0.13$~dex
for stars with metallicities [Fe/H] higher and lower
than -1.0, respectively. To avoid the uncertainties related to 
non-LTE effects for iron lines, we used everywhere the iron 
abundances determined solely from Fe\,II lines. Since the 
non-LTE corrections for magnesium in F-G dwarfs and subgiants 
are small and do not exceed 0.1~dex (Shimanskaya et~al. 2000),
they were not considered separately in our catalog. The relative 
magnesium abundances were derived from 1412 determinations in 
31 publications for 867 dwarfs and subgiants by using a 
three-pass iterative averaging procedure with the assignment 
of a weight to each primary source and to each individual
determination. The systematic shifts of all scales relative to 
the reduced mean scale were taken into account. Assigning the 
lowest weight to the least reliable determinations, this 
procedure yields final values close to those given by most 
sources without excluding any determination. The internal 
accuracies of the [Mg/Fe] values for metal-rich and 
metal-poor stars are $\varepsilon [Mg/Fe]= \pm 0.05$~dex and $\pm 0.07$~dex
respectively. 

The distances to the stars and their proper 
motions were calculated using data from currently available
high-accuracy catalogs,both orbital (Hipparcos –- Tycho) and 
ground-based (PPM-N, PPM-S, PPM-add) ones. We used the 
trigonometric parallaxes with errors of less than 25\,\% and,
in the absence of the latter the published photometric 
distances determined from uvbyb photometry. The radial 
velocities were taken mostly from the catalogs by Nidever 
et~al.(2002), Barbier-Brossat and Figon (2000), and Nordstrem 
et~al.(2004). In the absence of necessary data in these catalogs,
they were taken from other sources (see the references to Table~2 
in our previous paper (Borkova and Marsakov 2005)).

The components of the total space velocity ($U,V,W$) relative 
to the Sun were calculated for 844 stars with measured 
distances, proper motions, and radial velocities, where U is 
directed toward the Galactic anticenter V is directed in 
the sense of Galactic rotation, and W is directed toward 
the North Galactic Pole. At mean errors in the distances of 
15\,\% and a mean heliocentric distance of the stars from 
our sample of $\approx 60$~pc, the mean error in the space 
velocity components was $\approx \pm 2$~km~s$^{-1}$.

Based on the multicomponent model of the Galaxy containing a 
disk, a bulge, and an extended massive halo (Allen and 
Santillan 1991), we computed the Galactic orbital elements 
by simulating 30 revolutions of a star around the Galactic 
center. The Galactocentric distance of the Sun was assumed to
be 8.5~kpc, the rotational velocity of the Galaxy at the solar 
Galactocentric distance was 220~km~s$^{-1}$, and the velocity 
of the Sun with respect to the local standard of rest is 
$(U_\odot, V_\odot, W_\odot)=(-11, 14, 7.5)$~km~s$^{-1}$
(Ratnatunga et al.1989).

For our studies,we retained only 847 dwarfs and subgiants in 
the solar neighborhood with $T_{eff} > 5000$~K; in cooler stars,
the elemental abundances could be distorted. For 545 stars of 
our sample, we used the ages determined from theoretical 
isochrones in the paper by Nordstrem et~al.(2004). The absolute
magnitudes of the stars were found in this paper as in our paper 
mainly from the trigonometric parallaxes (from the photometric 
distances for some of the stars with large errors in the 
parallaxes), while the metallicities were derived from uvby 
photometry. A check shows that their photometric metallicities 
and temperatures are in good agreement with our spectroscopic 
data (the mean deviation is 
$\delta(\textrm{[Fe/H]}_{pf} - \textrm{[Fe/H]}_{sp}) = 0.07$~dex,
for the iron abundance at the dispersion of the deviations 
$\sigma_{\delta [Fe/H]} = 0.09$~dex; and
$\delta T_{ýôô}=45^\circ$ for the temperature at 
$\sigma_{\delta T_{ýôô}}=70^\circ$ i.~e., of the order of the 
above errors in the corresponding quantities in our catalog), 
which allows us to use the ages from this paper. The absolute 
values of the age errors given by the authors are rather large: 
the error is less than 1~Gyr only for $\approx 30\%$ of the 
stars, ranges from 1 to 2~Gyr for another $\approx 25\%$ of 
the stars, and is even larger up to $\approx 7$~Gyr for the 
remaining stars.

All stars of our sample fill the entire sky rather uniformly.
Nevertheless, it exhibits observational selection: the apparent 
magnitude depends on metallicity (see Fig.~7 in our previous 
paper (Borkova and Marsakov (2005)). On the one hand, it 
stemmed from the fact that mostly metal-richer stars lie near the
Galactic plane (where the Sun is located). On the other hand, 
observers knowingly use more sensitive instrumentation to 
increase the number of studied metal-poorer stars. As a result,
almost all of the stars with $[Fe/H ]\geq  -1.0$ were brighter 
than $V \approx 9^m$ while the bulk of the metal-poorer stars 
were fainter. However our results are not affected by this 
selection.

\section*{IDENTIFICATION OF STARS OF THE SUBSYSTEMS}

There is no single, necessary and sufficient criterion that 
would allow each star to be individually identified with absolute 
confidence by its belonging to a particular Galactic subsystem.
Only the mean values and dispersions of such key parameters of
the stellar subsystems as the ages, velocities, spatial locations,
Galactic orbital elements, metallicities, and relative abundances 
of certain chemical elements can be determined more or less 
accurately from reliably identified members of the subsystem. 
Any of the above parameters can act as a criterion for identifying 
the stars of various subsystems. For globular star clusters, the 
morphological structure of their horizontal branches and the 
mean periods of their RR~Lyrae stars (cf. Oosterhof's types) can 
also be attributed to these determining parameters. Choosing
any of these parameters, we get an opportunity to study the 
distributions of the subsystem's objects in other parameters. 
Such studies have revealed that no sharp boundaries exist between 
the subsystems and that stars with identical characteristics 
can belong to different subsystems. It would be appropriate to 
use complex criteria that include several parameters to identify 
the subsystem's objects more reliably.

The metallicity of stars is commonly used to separate the stars 
of the spherical Galactic subsystems from the diskones, since it 
remains, though rough, an age indicator. However in this paper we 
chose exclusively kinematic parameters as the initial criteria, 
since they shape the spatial distribution of objects that manifests 
itself in the morphological identification of several subsystems 
in the Galaxy. In this approach the possibility of observational 
selection that distorts the abundance distributions of various 
chemical elements in the stars of the subsystems, particularly
in the near-boundary regions,is ruled out. We separated the 
thick-disk stars from the halo using the following condition: 
$V_{res} = \sqrt{(U^2 + V^2 + W^2)} < 175$~km~s$^{-1}$, where 
$U, V, W$ and $V_{res}$ are the space velocity components and 
the total residual stellar velocity relative to the local 
standard of rest, respectively. We optimized the specific value 
of this criterion by minimizing the numbers of metal-rich 
($[Fe/H] >-1.0$) sample stars in the identified halo and 
metal-poor stars in the thick disk. The derived critical value
of the residual velocity for the selection of thick-disk stars 
from our sample is almost equal to that in the paper by 
Fuhrmann~(2000): $V_{res} = 180$~km~s$^{-1}$.
Chiappini et~al.~(1997) suggested a fairly informative kinematic 
index based on the same three space velocity components or 
more precisely on the Galactic orbital elements of stars 
calculated from them. This is $\sqrt{Z^2_{max}+4\cdot e^2}$, 
where $Z_{max}$ is the maximum distance of the orbital points 
from the Galactic plane, and $å$ is the orbital eccentricity 
of the star. Figure~1a shows the relation between these two 
kinematic indices. Even a distinct gap between the halo and 
the thick disk, which is indicated by the dotted line, is 
observed in the diagram (however it could also be the result 
of selection in the original lists of stars used to compile
the summary catalog).The identified subsystems are also 
naturally separated in the $[Fe/H]-\sqrt{Z^2_{max}+4\cdot e^2}$
plane, i.~e. composed of the two independent criteria
that we did not use,and form two clusters that can be 
arbitrarily separated (see the dotted line in Fig.~1b).
The stars lying in the immediate vicinity of this demarcation 
line occupy controversial positions here. However according 
to our kinematic criterion almost all of them fall into the 
thick disk as can be seen from the diagram. As a checkshows,
they fill rather uniformly the same regions in all kinematic 
diagrams of Fig.~1 as the remaining thick-disk stars. However
excluding these stars from the thick disk would introduce 
artificial chemical selection into both separated systems,
and a certain number of stars with an intermediate metallicity 
and with kinematics typical of the thick-disk stars would be 
attributed to the halo. Very metal-poor but rapidly revolving 
(around the Galactic center) stars are commonly called the 
metal-poor tail of the thick disk (Prochaska et~al.~2000). We
singled out 13 stars with $[Fe/H ]< -1.25$ from the 
kinematically selected thick-disk objects in order to
subsequently trace the correspondence of their properties to 
those of the subsystem's remaining stars. The metal-poor tail 
stars are denoted in Fig.~1 by the open triangles. Figure~1c 
shows the $\Theta$ - $\sqrt{Z^2_{max}+4\cdot e^2}$
diagram, where $\Theta$ is the circular velocity of the star
around the Galactic center. We see from this figure that the 
halo and thick-disk stars (except several stars) in this 
diagram are separated by a clear gap, while the metal-poor 
tail stars fill uniformly the region occupied by all of the 
remaining thick-disk stars. The location of the gap is 
highlighted in the diagram by the inclined dotted line. The 
distribution of stars of the identified subsystems in 
Toomre's diagram is shown in Fig.~1d. We emphasize that 
any criterion is purely statistical and the membership of 
each specific star remains doubtful. For example, according 
to the residual velocity criterion, the star HD~97916 falls 
into the thick disk but its high value of $Z_{max} > 4$~kpc,
which is atypical of the stars of this subsystem (see Fig.~6 
below), suggests that it can also belong to the halo with 
equal success (see Fig.~1). Based on the $3\sigma$ criterion 
applied to the parameter $Z_{max}$, we excluded this star 
from the subsequent analysis among the thick-disk stars.

The thin-disk stars are known to have low residual velocities 
with respect to the local standard of rest and nearly circular 
orbits all points of which do not rise high above the Galactic 
plane. Therefore, for reliability we identified this subsystem 
using simultaneously two kinematic criteria: 
$V_{res} <85$~km~s$^{-1}$ (see also Furmann 1998) and 
$\sqrt{Z^2_{max}+4\cdot e^2} < 1.05$. Higher-velocity stars 
were included in the thick disk. In this case, the values of 
the criteria were chosen to minimize the number of stars 
with a high relative abundance of magnesium in the thin 
disk and with its low abundance in the thick disk. It turned 
out that we could not get rid of the overlap between the
abundance ranges of various chemical elements in the stars 
of the subsystems being separated by any combination of any 
kinematic parameters. However the 
$V_{res}$--$\sqrt{Z^2_{max}+4\cdot e^2}$ diagram (Fig.~1a) 
reveals even sparseness of stars between the subsystems. 
Figures~1a and 1d show that we cannot reliably determine
which subsystem a star belongs to from any single velocity 
component. Indeed, according to the adopted criteria, a star 
in the thick disk can have any arbitrarily low velocity 
component, but the other velocity components (or only one 
component) will necessarily be appreciably higher than 
those for the disk stars.

If we follow the kinematic criterion then the metallicity is 
represented by not quite a proper criterion. Indeed, a 
significant overlap of the metallicity ranges between the 
thick and thin disks is seen from Fig.~1b. Consequently the 
criterion based on the position of the dip found in the 
metallicity function of field stars near $[Fe/H ]= -0.5$ 
(see, e.~g. Marsakov and Suchkov 1977), which is commonly 
used to separate the stars of the subsystems, can be used 
only when there are no data on the total space velocities of
the stars. The situation with the age is also difficult. Based 
on the accurately determined ages of only seven subgiants, 
Bernkopf et~al.~(2001) concluded that the thin disk began 
to form $\approx 9$~Gyr ago, while the thick-disk stars are 
definitely older than 12~Gyr. Nevertheless, the ages,
unfortunately cannot yet be used as a reliable criterion 
for separating the two disk subsystems, because the 
individual ages of most stars are determined with a low 
accuracy. Therefore in this paper we use the ages only to 
reveal statistical trends.

\section*{RELATION BETWEEN THE IRON AND MAGNESIUM ABUNDANCES}

Let us consider the properties of the stars of the thick disk
that we identified. Elucidating the question of whether 
there is a trend in the relative magnesium abundance with 
metallicity inside the thick diskis of greatest importance here.
Figure~2a shows the [Mg/Fe]-–[Fe/H] diagram for all stars of our 
catalog where the identified thick-disk stars are highlighted
by the triangles. To checkthe results using more reliable data,
exactly the same diagram, but only for the stars in which the 
magnesium abundances were averaged over several sources and with 
the total weight of the mean value larger than unity is shown
in Fig.~2b. Figure~2c shows the stars of only the thick disk and 
plots the median curve that roughly fits the pattern of [Mg/Fe] 
variation with increasing metallicity. Unfortunately this 
dependence cannot be constructed mathematically rigorously; 
therefore, we drew it by eye halfway between the upper and
lower envelopes of the diagram. We see from the figure that an 
appreciable number of stars with thick-disk kinematics have a 
chemical composition typical of the thin-disk stars, with the 
magnesium trend with increasing metallicity for these stars 
being very similar to that in the thin disk. Most stars 
(9 of the 13) of the metal-poor thick-disk tail (open triangles)
show approximately the same magnesium abundance, 
$\langle[Mg/Fe]\rangle = 0.40\pm0.01$ but some of them (four stars) at 
metallicity $[Fe/H ]< -1.7$ exhibit a very low abundance 
ratio, $[Mg/Fe ]=(-0.3 \div 0.3)$~dex which is atypical of 
the stars formed from matter with the same chemical evolution.
Starting from $[Fe/H ]\approx -1.0$ (the vertical dotted line 
in the figure), the tendency for the relative magnesium 
abundance to decrease with increasing metallicity begins to 
show up increasingly clearly in the thick disk. Even if we 
restrict our analysis to the magnesium abundance range 
$[Mg/Fe]>0 .25$ the correlation coefficient turns out to be 
larger than zero beyond the error limits. However even when 
the boundary is lowered to $[Mg/Fe]=0.20$, i.~e. to the
upper limit for the overwhelming majority of thin-disk stars,
the correlation coefficient becomes equal to $r =0.4 \pm 0.1$ 
(see the inclined straight lines in Figs.~2a and 2b). If, 
however, we include metal-poorer thick-disk star up to 
$[Fe/H ]= -1.25$ then the slopes of the regression lines in 
both diagrams will decrease slightly. However the figure 
shows a considerable number of thick-disk stars with even 
lower magnesium abundances, and the correlation coefficient
increases to $r =0.65 \pm 0.05$ for all of the thick-disk
stars with metallicities $[Fe/H ]> -1.0$. This increase
in correlation is attributable mainly to a sudden decrease 
in the mean magnesium abundance starting from 
$[Fe/H ]\approx -0.6$ which showed up as a sharp bend of the 
median sequence in Fig.~2c. The sequence again becomes flatter 
starting from $[Fe/H ]\approx -0.35$. Thus the data from 
our catalog suggest that the [Mg/Fe] ratio decreases with 
increasing [Fe/H ] in the thick disk and that the explosions 
of the first SNe\,Ia began when the mean metallicity of the 
interstellar medium in the Galaxy reached $[Fe/H]\approx -1.0$.
Howeverthe explosions of SNe\,Ia began later than the mass 
star formation in the thick disk which occurred (as can be 
seen from Fig.~2) even at $[Fe/H ]\approx -1.25$. These results 
confirm the conclusion about a fairly long duration of the 
star formation in the subsystem that was drawn by 
Prochaska et~al.~(2000) from the trends in three 
other $\alpha$-elements with metallicity detected by the 
authors. Prochaska et~al.~(2000) argue that such a long 
time scale is possible only in the thick-disk formation 
scenario during dissipational collapse. The formation of 
metal-poor tail stars in this scenario can be explained by 
the fact that a number of stars with kinematics 
characteristic of the thick-disk stars were formed from 
isolated clouds of interstellar matter less enriched with 
heavy elements. The existence of stars with a low 
magnesium abundance ($[Mg/Fe ]<0.2$~dex) in the metal-poor 
tail at low heavy-element abundances most likely implies
that these clouds were in weak contact with the bulk of 
the protogalactic cloud in which the matter was mixed 
fairly well.

A further argument for the latter suggestion is the result 
by Navarro et~al.~(2004). They provide evidence that the 
stars of the long known Arcturus moving group (see Eggen~1998)
may well originate from a massive satellite disrupted by 
the Galactic tidal forces at an early formation phase of the 
Galaxy. We selected stars with close angular momenta from the
original catalog according to the criteria in the paper
by Navarro et~al.(2004): $(85 < \Theta < 130)$, 
$|W| < 50$~êì~ñ$^{-1}$ and $|\Pi| < 150$~êì~ñ$^{-1}$, where 
$\Pi$, $\Theta$, and W are the space velocity components of 
the stars with respect to the local standard of rest in the 
cylindrical coordinate system. It turned out that 15 of the 
17 stars selected in this way belong to our thick disk and 
3 of them belong to its metal-poor tail (two stars fell into
the protodisk halo). Figure~2 shows the [Mg/Fe]–-[Fe/H] diagram 
for the thick-disk stars (triangles); the circles denote the 
stars of the Arcturus group. The considerably smaller spread 
in [Mg/Fe] ratios than that for the thick-disk stars at given 
metallicity for Arcturus group candidates argues that these 
stars may well have a common origin. Thus we see that at least 
some of the stars with thick-disk kinematics can be extragalactic 
in origin, i.~e. be formed from matter with a different history 
of chemical evolution. 

All of the currently existing thick-disk formation scenarios 
can be divided into two classes:(1) the collapse of a single 
protogalactic cloud; and (2) the result of the interaction 
between the early Galaxy and nearby satellite galaxies. The 
behavior of the subsystem's stars in different scenarios must 
differ. Thus, for example, according to the theoretically 
modeled first scenario a vertical gradient in chemical 
composition must be formed in the thick disk through the collapse 
of the protogalactic cloud (Burkert et~al.~1992). However free 
collapse takes a very short time, $\approx 400$~Myr while the 
relative abundances of chemical elements in the interstellar 
medium can change appreciably only on much longer time scales.
The above evidence for the existence of SNe\,Ia traces in the 
chemical composition of the subsystem's stars suggests that 
the overwhelming majority of stars in it were formed after 
the mass explosions of supernovae of this type began, i.~e.
no earlier than $1–2$~Gyr (see, e.~g. Matteucci~2001) after 
the initial star formation in the protogalactic cloud. By 
this time according to the hypothesis of rapid collapse, all 
of the interstellar matter in the protogalaxy had already 
concentrated near the plane. This contradiction is naturally 
circumvented in the theoretical model where as the result of 
a close encounter with a fairly massive satellite galaxy
the already formed stars of the primordial thin disk acquire 
a significant acceleration mainly perpendicular to the 
Galactic plane, producing the thick disk (Kroupa~2002). In 
addition, at all evolutionary phases, the tidal forces of 
the early Galaxy could disrupt dwarf satellite galaxies, 
capturing their stars, and as predicted by the corresponding 
theoretical model (Quinn et~al.~1993), it is then unlikely 
that a vertical gradient in chemical composition will arise 
in its thick disk. The latter assumption is supported by 
the possible extragalactic origin of the Arcturus moving group,
almost all the stars of which belong to the thick disk.

To concretize the model, let us analyze the relation between 
the chemical composition and other parameters of the 
subsystem's stars.

\section*{RELATION OF THE IRON AND MAGNESIUM ABUNDANCES TO ORBITAL
ELEMENTS AND AGE}

Having reconstructed the Galactic orbits of the stars, we can 
trace the variations in metallicity and relative magnesium 
abundance in the subsystem with maximum distance of the 
stars from the Galactic center and plane. Figure~3a shows 
the $Z_{max} - [Fe/H]$ diagram for all thick-disk stars. The
thirteen metal-poor tail stars with $[Fe/H ]< -1.25$
are highlighted by the open triangles and are not used to 
determine the gradients. The regression line constructed for 
the remaining stars exhibits a fairly large negative vertical 
metallicity gradient: grad$_Z$[Fe/H]=$(-0.13 \pm 0.04)$~kpc$^{-1}$ 
at $r = -0.27 \pm 0.08$. The result is stable, since even 
eliminating the four most distant stars lying farther than 
2~kpc do not change the slope of the regression line at
all. The correlation is highly significant, since the probability 
of random occurrence of the correlation is $P_N \ll 1\%$. 
The vertical gradient in the thick disk derived from the nearest 
stars is equal, within the error limits, to the value of 
$-0.2 \pm 0.1$~kpc$^{-1}$ obtained from RR~Lyrae variable field 
stars (see Borkova and Marsakov~2002). The thick-disk stars 
also exhibit a vertical gradient in relative magnesium 
abundance (see Fig.~3b): grad$_Z$[Mg/Fe]=(0.06$\pm$0.02)~kpc$^{-1}$
at $r =0.33 \pm 0.08$ and at a probability of random 
occurrence of the correlation $P_N \ll 1\%$; i.~e., the 
correlation can be recognized to be highly significant.
Here,removing the four most distant stars changes neither 
the correlation coefficient nor the gradient. Note that the 
correlations in Figs.~3a and 3b stem exclusively from the 
existence of two groups of points that are separated by a 
distinct deficit of stars near $Z_{max}\approx 0.5$~kpc; 
it is not quite clear why it arose (see also the $Z_{max}$ 
distribution of stars in Fig.~6a). There are no correlations 
inside each group but they differ in both mean magnesium 
abundance and metallicity. Therefore, if we used a more 
stringent criterion and did not include low-orbit stars 
in the subsystem (in Fig.~1 this roughly corresponds to 
the satisfaction of the condition 
$\sqrt{Z^2_{max}+4\cdot e^2} > 0.8$ for the thickdisk),
then the two vertical gradients in it would disappear.
However the high orbital eccentricities 
($\langle e\rangle = 0.33 \pm 0.01$), residual velocities 
($\langle V_{res}\rangle = 97 \pm 2$~km~s$^{-1}$),
and ages ($\langle t\rangle = 9.7 \pm 0.9$~Gyr) and the low azimuthal
space velocity ($\langle\Theta\rangle = 190 \pm 6$~km~s$^{-1}$)
even at low magnesium abundance ($[Mg/Fe]=0.21 \pm 0.02$) in
the excluded stars do not allow them to be attributed to the 
thin disk. Therefore some of the authors (e.~g., Fuhrmann~1998;
Bensby et~al.~2003) did not include them in any of the 
disk subsystems, calling them transitional stars. Thus the 
question as to whether a vertical gradient in chemical 
composition exists in the thick disk will be solved only 
after it will be established which subsystem this group 
of stars belongs to.

The radial metallicity gradient in the thick disk (Fig.~3c) 
was found to be positive rather than negative, but almost 
indistinguishable from zero outside the error limits 
($r =0.18 \pm 0.09$). Since the probability of random 
occurrence of this correlation is $P_N = 5~\%$ it cannot 
be recognized to be statistically significant. The RR~Lyrae 
variable stars in the thick disk reveal no radial metallicity 
gradient at all (Borkova and Marsakov~2002). In contrast, 
the radial gradient in relative magnesium abundance is not 
only nonzero outside the $3\sigma$ limits, but also is 
negative, instead of the expected increase in the relative 
magnesium abundance with increasing apogalactic orbital 
radius of the stars: 
grad$_{R_a}$[Mg/Fe]=(-0.026$\pm$0.007)~kpc$^{-1}$ at 
$r = -0.34 \pm 0.08$ and at a probability of random 
occurrence of the same correlation $P_N \ll 1~\%$ (see Fig.~3d).
Note that the corresponding radial gradients and correlation 
coefficients in Figs.~3e and 3f were found to be almost the 
same when using the mean orbital radii of the stars. The 
latter are believed to be less subject to variations 
during the stellar lifetime and therefore, are commonly used 
to find the initial radial metallicity gradients. Remarkably 
eliminating the stars with $\sqrt{Z^2_{max}+4\cdot e^2} < 0.8$ 
from the subsystem has virtually no effect on the results,
since the stars with any orbital radii are eliminated more 
or less equiprobably. We clearly see from the diagrams that 
the density of stars with a high magnesium abundance increases 
continuously with decreasing galactocentric distance and the 
overwhelming majority of them are found near the solar 
orbital radius, while the density of magnesium-poor stars in 
the diagrams is almost independent of both $Ra$ and $\langle R\rangle$.
Therefore,even if we reject the seven most distant
stars in the diagrams (with $Ra >13$~kpc in Fig.~3d and with 
$\langle R\rangle \ > 10$~kpc in Fig.~3f), both the slopes
of the regression lines and the correlation coefficients 
will only increase. The relative numbers of stars with different 
metallicities in the [Fe/H]--Ra (Fig.~3c) and
$[Fe/H]-\langle R\rangle$ (Fig.~3e), where positive 
correlations are observed instead of the expected negative 
correlation behave in a similar way. This gradient inversion
is most likely caused by the selection effect related to the 
predominantly smaller perigalactic orbital radii and larger 
eccentricities for the metal-poorer, and simultaneously 
magnesium-richer stars in the solar neighborhood (see the 
correlations in the diagrams of Fig.~4). Indeed the probability 
of finding a star near its apogalactic orbital radius is 
highest. Therefore among the thick-disk stars with fairly 
high orbital eccentricities ($\langle e \rangle = 0.34$ at 
$\sigma_e = 0.13$), we will see a larger fraction of slow 
stars near the apogalactic radius equal to the solar orbital 
radius than stars with large $Ra$ and $\langle R\rangle$ 
passing rapidly by the Sun. At the same time, the relative 
velocities for the stars with the perigalactic radii equal 
to the solar orbital radius are low; therefore, among the 
stars with large $Ra$ and $\langle R\rangle$ in the solar 
neighborhood, the fraction of such stars is larger. Since 
both the magnesium abundance ($r = -0.44 \pm 0.07$ at 
$P_N \ll 1\%$), and to a small degree, the metallicity 
($r =0.18 \pm 0.09$ at $P_N \ll 5\%$) in the thick disk
correlate with $Rp$ (see Figs.~4a and 4c, respectively),
magnesium-poorer, and simultaneously metal-poorer stars 
will prevail at large $Ra$. The correlation between magnesium 
abundance and eccentricity ($r =0.34 \pm 0.08$ at $P_N \ll 1\%$)
in Fig.~4b also contributes to the enhancement of this 
selection. In Fig.~4d metallicity does not correlate with
eccentricity; nevertheless, its sign is negative: 
$r =-0.11 \pm 0.09$ at $P_N \ll 28\%$. We emphasize that 
both correlations with the magnesium abundance should
be recognized to be highly significant and eliminating 
the low-orbit stars from the subsystem does not change the 
result. Thus, the usually declared absence of a radial 
metallicity gradient and the existence of the negative 
(inverse) radial magnesium abundance gradient in the thick 
disk found above can be largely attributed to the selection 
related to the observations of stars in a limited region of 
Galactic space.

The distributions of the stars that are currently in
the local volume of the Galaxy in orbital eccentricity
and maximum distance from the Galactic plane are
less dependent on the observer's location, i.~e. the
Sun than their distributions in maximum and minimum Galactocentric 
orbital radii. Therefore, these orbital elements may be 
expected to also correlate with stellar age. We see from Fig.~5 
that the two parameters indeed exhibit a correlation with age.
However the $e(t)$ correlation is actually significant (the 
correlation coefficient is $r =0.35 \pm 0.10$ at $P_N \ll 2\%$),
while the (t -- $Z_{max}$) diagram exhibits no correlation
but it clearly shows the absence of large distances from the 
Galactic plane among the young stars (see the empty upper 
left corner in Fig.~5b separated by the dotted line). The 
magnesium abundance also correlates with age 
($r =0.27 \pm 0.11$ at $P_N \ll 2\%$ i.~e., the correlation 
should still be recognized to be significant), while the 
metallicity does not correlate with age at all (see Figs.~5c 
and 5d,respectively). Eliminating the low-orbit stars 
changes nothing. Let us [Mg/Fe]--[Fe/H] consider the t --[Mg/Fe]
diagram in more detail. According to Fuhrmann~(2000), there can 
be no stars younger than $\approx 10$~Gyr in the thick disk 
but they are present in our diagram. The larger triangles 
in the diagram denote the stars with age errors of less
2~Gyr. We see that the number of stars decreased by a factor 
of 2 but the overall structure of the diagram did not change 
and the correlation was retained. The stars in the diagram 
are naturally separated into two groups by the dotted line 
drawn almost to the regression line. The groups of different 
ages turned out to differ in both mean space velocity and 
galactic orbital elements. In particular their mean residual 
velocities $\langle V_{res}\rangle$=(95$\pm$3)~km~s$^{-1}$ and
(116$\pm$3)~km~s$^{-1}$, their dispersions  
$\langle\sigma_{V_{res}}\rangle$=(13$\pm$2)~km~s$^{-1}$ and 
$25 \pm 2$~km~s$^{-1}$, azimuthal velocities 
$\langle\Theta\rangle$=(200$\pm$7)~km~s$^{-1}$ and 
$156 \pm 4$~km~s$^{-1}$, orbital eccentricities 
$\langle e \rangle$=(0.31$\pm$0.01) and $0.36 \pm 0.01$, and
the maximum distances of orbital points from the Galactic plane 
$\langle Z_{max}\rangle$=(0.55$\pm$0.10)~kpc and $0.87 \pm 0.07$~kpc,
respectively for the young and old groups. A subgroup of six 
stars with a high magnesium abundance (the metal-poor tail 
stars were not included) stands somewhat apart from the 
stars with young ages. The mean values of its kinematic 
parameters listed above are almost equal to those for the main
subgroup of young stars. The subgroups differ only in magnesium 
abundance and metallicity: for the subgroup of six stars,
$[Fe/H ]= -0.74 \pm 0.10$~dex is almost equal to the 
metallicity of the old group $-0.63 \pm 0.03$~dex while for 
the main young subgroup, it is $-0.35 \pm 0.05$~dex. The
fact that 8 of the 24 stars in the young magnesium-poor 
subgroup have a circular velocity $\Theta > 220$~km~s$^{-1}$ 
and $Ra >12$~kpc was unperpendicularly expected. There are 
virtually no such rapidly revolving stars in the old group.
Judging by their characteristics, these stars could well be 
attributed to the thin disk if it were not for the larger 
$Z_{max} >1$~kpc for half of them, and $e>0.3$ which is a 
typical of the thin-and disks tars, for the remaining stars.
Thus, it seems that the thick disk consists of two discrete 
populations that differ in ages, elemental abundances, angular
momenta, space velocity dispersions, and orbital elements. 
However the belonging of the young group to the thick disk requires 
additional proofs.

The scale height is one of the most important parameters of 
the subsystem. It characterizes the subsystem's thickness and 
is equal to the height above the Galactic plane at which the 
density of its objects decreases by a factor of $e$. This 
parameter can also be determined from nearby stars by 
constructing their $Z_{max}$ distribution and fitting it by an
exponential law. The following should be taken into account:
first, the kinematic selection, since the probability of 
detecting a star at a given time near the Sun decreases with 
its increasing relative velocity; and, second, we must not 
just include each star in the histogram with its $Z_{max}$ 
but as it were, "spread" it over the orbit from $-Z_{max}$ to 
$+Z_{max}$ proportional to the probability density of finding 
it at diferent $Z$. This is because all stars of the subsystem 
cannot be simultaneously at the maximum distances from the
Galactic plane. To allow for the kinematic selection, we 
assigned a weight to each star proportional to its residual 
velocity with respect to the local standard of rest, $V_{res}$.
The initial $Z_{max}$ histogram and the reconstructed actual 
$Z$ distribution of thick-disk stars are shown in Fig.~6. 
The solid curve in Fig.~6b represents a function of the 
form $n(Z)=C \cdot \exp(-Z/Z_0)$ where $Z_0$ is the scale 
height (for more details on the procedure for reconstructing 
the height distribution of stars, see Marsakov and Shevelev~(1995)).
The scale height for the thick disk was found from the stars of 
our sample to be $Z_0 =0.6 \pm 0.1$~kpc, where the uncertainty 
was estimated from the maximum and minimum values of $Z_0$ 
obtained by eliminating the extreme columns of the histogram.
The derived scale height is slightly smaller than its value
that we determined also from the RR~Lyare variable field stars:
$Z_0 =0.74 \pm 0.05$~kpc (see Borkova and Marsakov~2002),
while the latter matches the estimates of other authors (in 
particular Robin et~al.~(1996) gave a value averaged over 
several source of $Z_0 =0.76 \pm 0.05$~kpc). The size of the 
sample used must be increased significantly to reliably 
estimate the scale height for the subsystem. Note also that 
the scale height slightly increases if the low-orbit stars
are eliminated from the subsystem.

The mean velocity components and their dispersions calculated 
from all 133 thick-disk stars of our sample are: 
$\langle U\rangle$=(4$\pm$7)~km~s$^{-1}$ 
$\sigma_U$=(80$\pm$5)~km~s$^{-1}$;
$\langle V\rangle$=(-49$\pm$4)~km~s$^{-1}$ 
$\sigma_V$=(46$\pm$3)~km~s$^{-1}$;
$\langle W\rangle$=(2$\pm$4)~km~s$^{-1}$ 
$\sigma_W$=(45$\pm$3)~km~s$^{-1}$.
They are in satisfactory agreement with the determinations
of other authors even in the case of a radical difference 
between the initial prerequisites in forming the samples of 
thick-disk stars. For example, Soubiran et~al.~(2003) obtained 
($\sigma_U$, $\sigma_V$, $\sigma_W$)=(63$\pm$6, 39$\pm$4, 
39$\pm$4)~km~s$^{-1}$ for a sample of $\approx 400$ thick-disk
red giants lying toward the North Galactic Pole. On the other 
hand, Gilmore and Wyse~(1987) give the following velocity 
dispersions for the thick-disk field stars: $(70,50,45)$~km~s$^{-1}$.

\section*{THE FORMATION TIME SCALE OF THE THICK DISK}

Thus the trends that we found here strongly suggest that the 
formation time scale of the thick disk was rather long and in no 
way fits into the scenario of a rapidly collapsing protogalaxy (Eggen 
et~al.~1962). The formation of the first stars in it began long 
before the onset of mass type-Ia supernovae explosions when the 
metallicity had not yet reached a turning value of $[Fe/H] \approx -1.0$.
Since the evolution time scale of close binary stars that end their lives
with SNe\,Ia explosions is $\approx(1 - 2)$~Gyr the formation
of the subsystem began before the Galaxy reached this age. By this 
time, the young Galaxy already had an almost formed spherical 
subsystem. By studying the variations in the abundances of 
r-- and s--elements in thick-disk stars with magnesium 
abundance $[Mg/Fe]>0.25$ Mashonkina et~al.~(2003) showed that the 
overwhelming majority of thick-disk stars were formed within 
1.1 to 1.6~Gyr after the protogalactic cloud began to collapse.
However the last stars in the thick disk turned out to have been
born after the formation of a younger subsystem, the thin disk, began.
This is evidenced by our largely selection-free sample that contains 
an appreciable number of stars with thick-disk kinematics, but lying
in the ranges of metallicities and relative magnesium abundances 
extending to their solar values. Indeed, according to the isochrone 
ages, these stars are no older than the Sun. The angular momentum 
difference between the halo and the thick disk suggests the absence 
of a smooth transition between these two subsystems 
(Wyse and Gilmore~1992). A long delay of star formation after the 
mass supernova explosions, an active phase in the evolution of 
the Galaxy (see Berman and Suchkov~1991), could be responsible
for this difference. This explanation simultaneously resolves the 
contradiction with the short time scale of free collapse of the 
protogalactic cloud ($\approx 0.4$~Gyr), and the age at which the 
mass formation of thick-disk stars begins could prove to be even 
older than 1~Gyr. In other words the delay of star formation 
lengthens the time frames of the natural processes of collapse
and chemical enrichment of the interstellar medium.

The vertical gradients in metallicity and magnesium abundance 
that we found as well as the correlations between the magnesium 
abundances (and, possibly also metallicities) and the perigalactic 
distances and the magnesium abundances and the orbital 
eccentricities naturally fit into this picture. However the
question arises as to whether the stars with thick-disk 
kinematics, but lying in low orbits belong to the subsystem,
since, if they do not belong to the thick disk there are no 
vertical gradients in it. In this case, the model of the 
subsystem's formation through the interaction of the Early 
Galaxy with its satellites becomes of current interest. The 
stars of the so-called metal-poor thick-disk tail do not 
follow the patterns related to [Fe/H] and exhibited by all 
the remaining stars of the subsystem. In addition, they 
show an anomalously large spread in [Mg/Fe]. Therefore,
we believe that they were most likely formed inside isolated 
interstellar clouds that were enriched with chemical elements 
according to scenarios different from the evolution of the 
actively mixing main body of the interstellar matter of a 
single protogalactic cloud. Further studies and data on the 
abundances of other chemical elements in the subsystem's 
stars are required to refine the thick-disk formation time 
scale and model.

In the next papers,we are going to consider the properties of 
stars in other Galactic subsystems and to further supplement 
the existing database of stars in the solar neighborhood with 
published determinations of the abundances of various chemical 
elements.

{\bf Acknowledgements}: 
We are grateful to the anonymous referee for valuable remarks 
that forces us to present our results in a more reasoned way.

\newpage

\begin{figure*}
\centering
\includegraphics[angle=0,width=0.99\textwidth,clip]{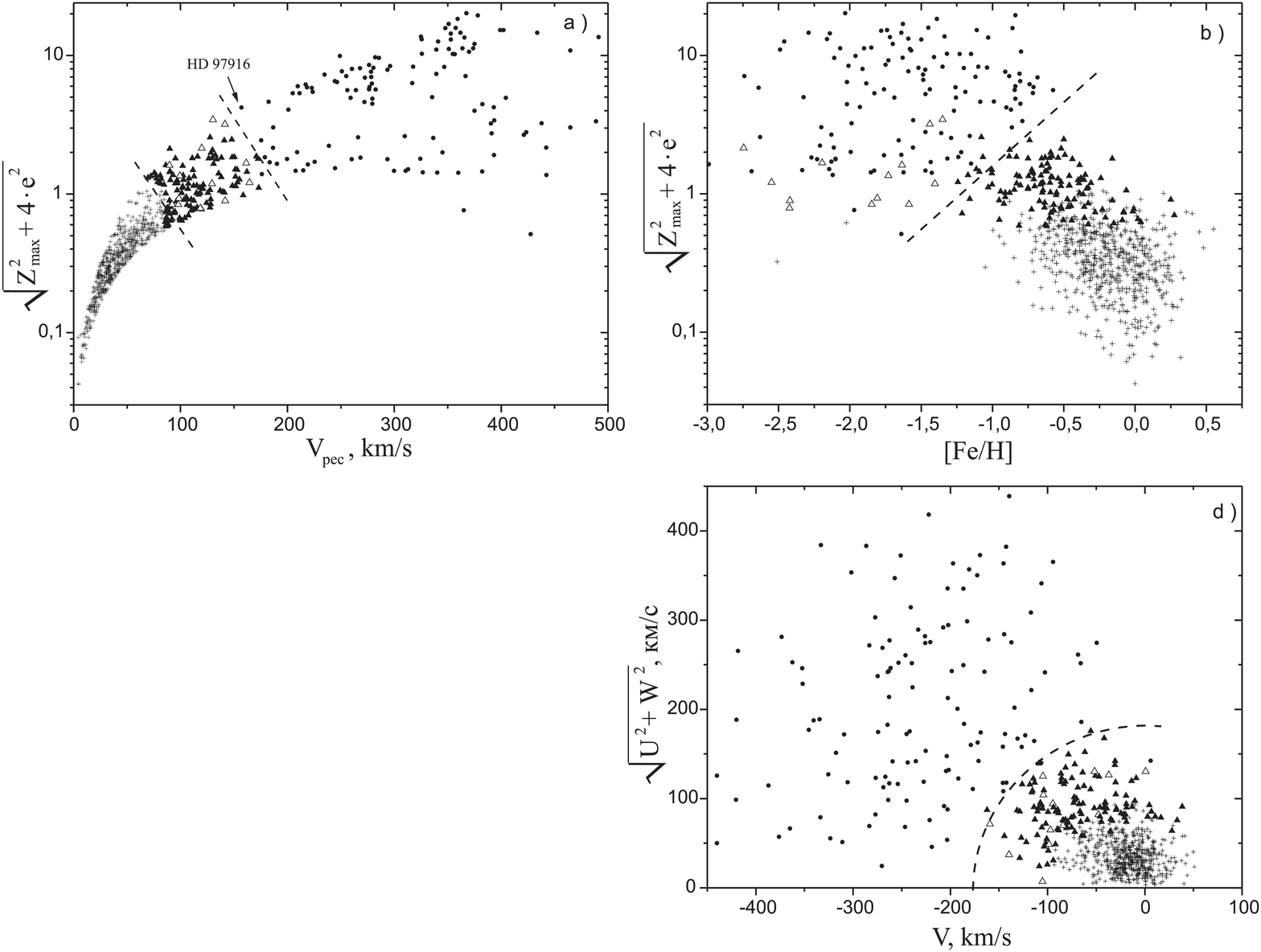}
\caption{Relations between $\sqrt{Z^2_{max}+4\cdot e^2}$ and the residual stellar 
velocities with respect to the local standard of rest (a) metallicities (b),
circular velocities (c) and between the residual stellar velocity components (d):
crosses -- thin-disk stars; filled triagles -- thick-disk stars; 
open triagles -- stars of the metal-poor thick-disk tail;
filled circles -- halostars; and dotted lines -- demarcation lines drawn by eye.
}
\label{fig1}
\end{figure*}

\newpage

\begin{figure*}
\centering
\includegraphics[angle=0,width=0.84\textwidth,clip]{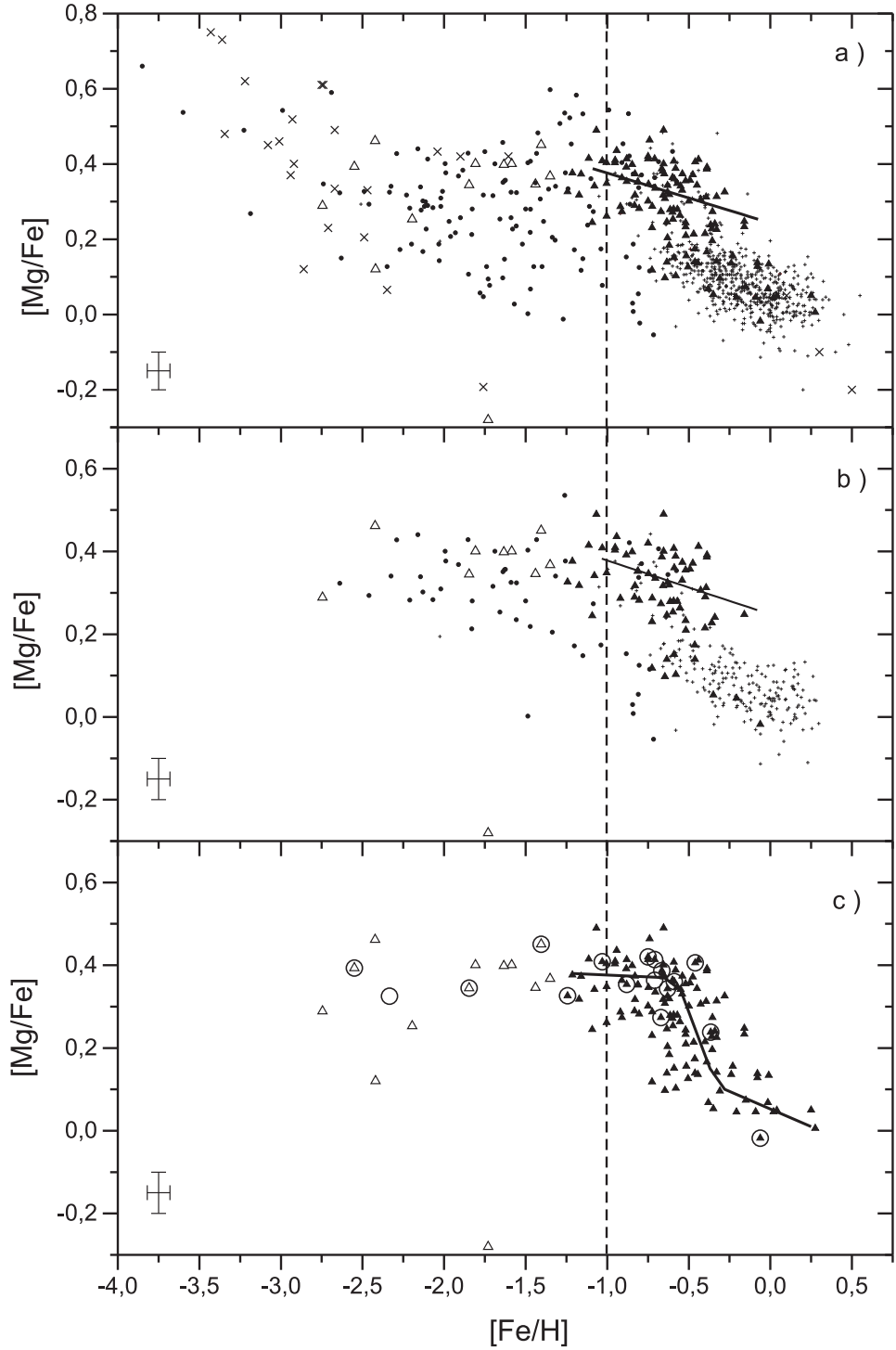}
\caption{Relations between metallicity and relative magnesium abundance (a)
for all stars of our catalog, (b) for the stars in which the magnesium
abundances were averaged over several sources, and (c) only for the 
thick-disk stars. The notation is the same as that in Fig.~1; the 
crosses denote the stars with unmeasured radial velocities. The
vertical dotted lines indicate the point at which the knee of the 
([Mg/Fe]-–[Fe/H]) relation begins, and the inclined straight lines 
represent the regression lines for metal-rich and, simultaneously 
magnesium-rich thick-disk stars; the broken curve represents the 
median line drawn by eye. The open circles in the panel denote the 
stars of the Arcturus moving group. The error bars are shown.}
\label{fig2}
\end{figure*}

\newpage

\begin{figure*}
\centering
\includegraphics[angle=0,width=0.99\textwidth,clip]{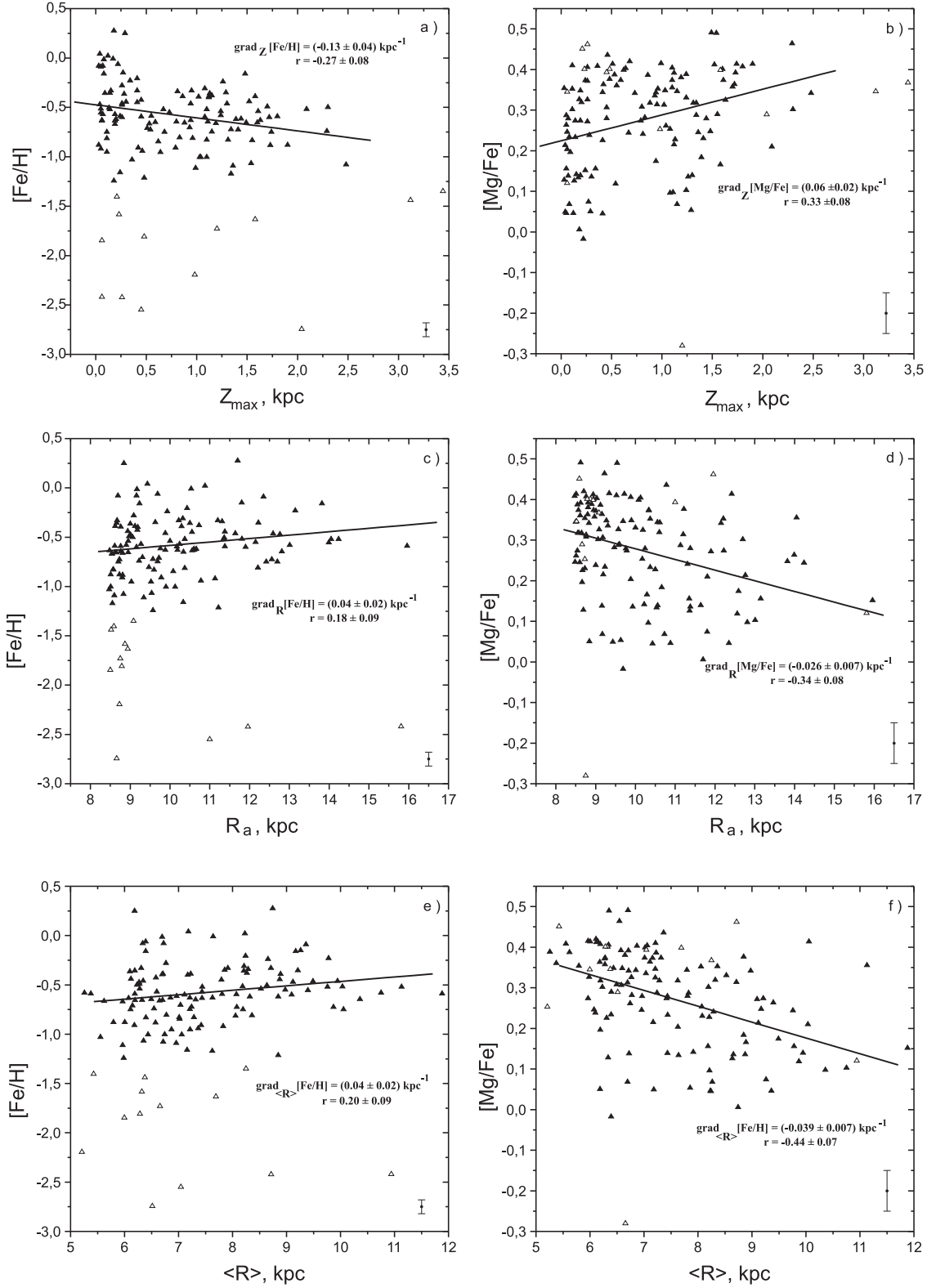}
\caption{Relations of the metallicity and relative magnesium 
abundance in the atmospheres of thick-disk stars to their 
maximum distance fromthe Galactic plane (a), (b), to the 
apogalactic distance (c), (d), and to the mean orbital radius 
(e), (f). The notation is the same as that in Fig.~1. The 
solid lines represent the regression lines for the thick-disk 
stars. The gradients and correlation coefficients are shown. 
The error bars are given.}
\label{fig3}
\end{figure*}

\newpage

\begin{figure*}
\centering
\includegraphics[angle=0,width=0.99\textwidth,clip]{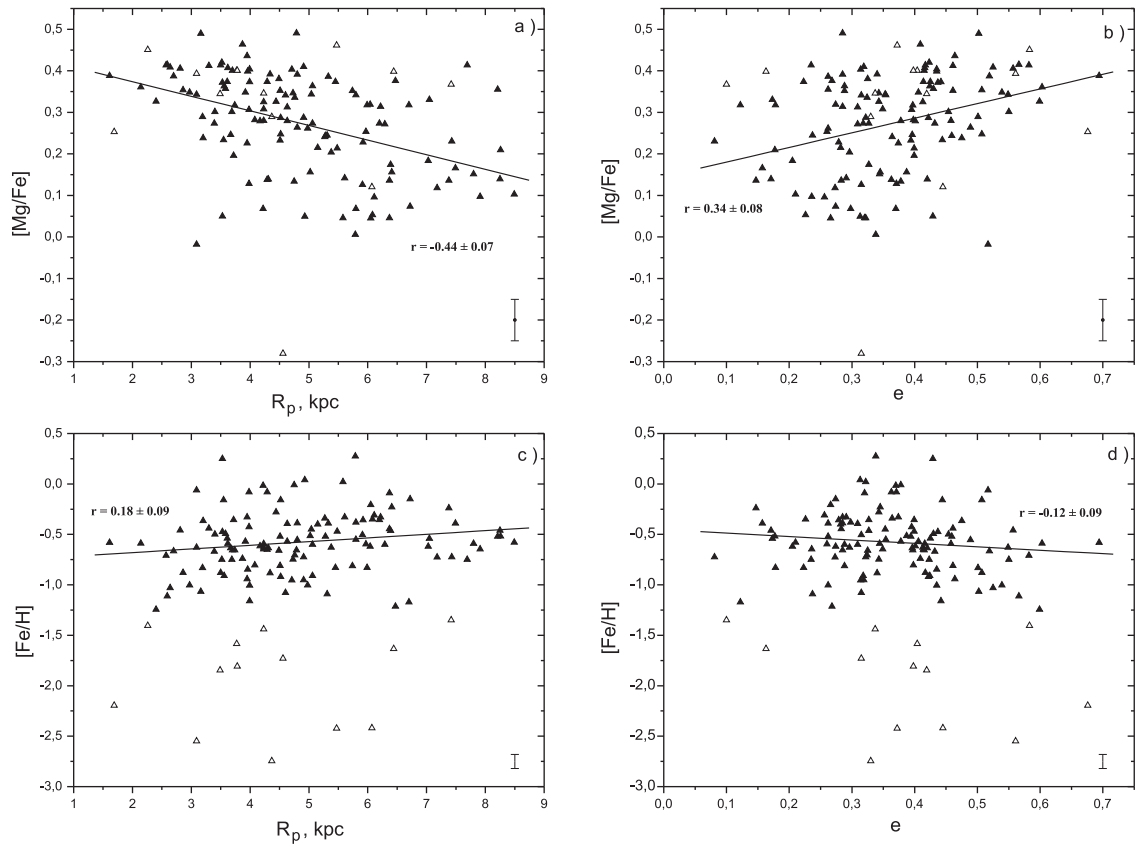}
\caption{Relations of the metallicity and relative magnesium 
abundances in the thick-disk stars to their perigalactic distances
(a), (c) and orbital eccentricities (b), (d). The correlation 
coefficients are shown. The error bars are given.}
\label{fig4}
\end{figure*}

\newpage

\begin{figure*}
\centering
\includegraphics[angle=0,width=0.99\textwidth,clip]{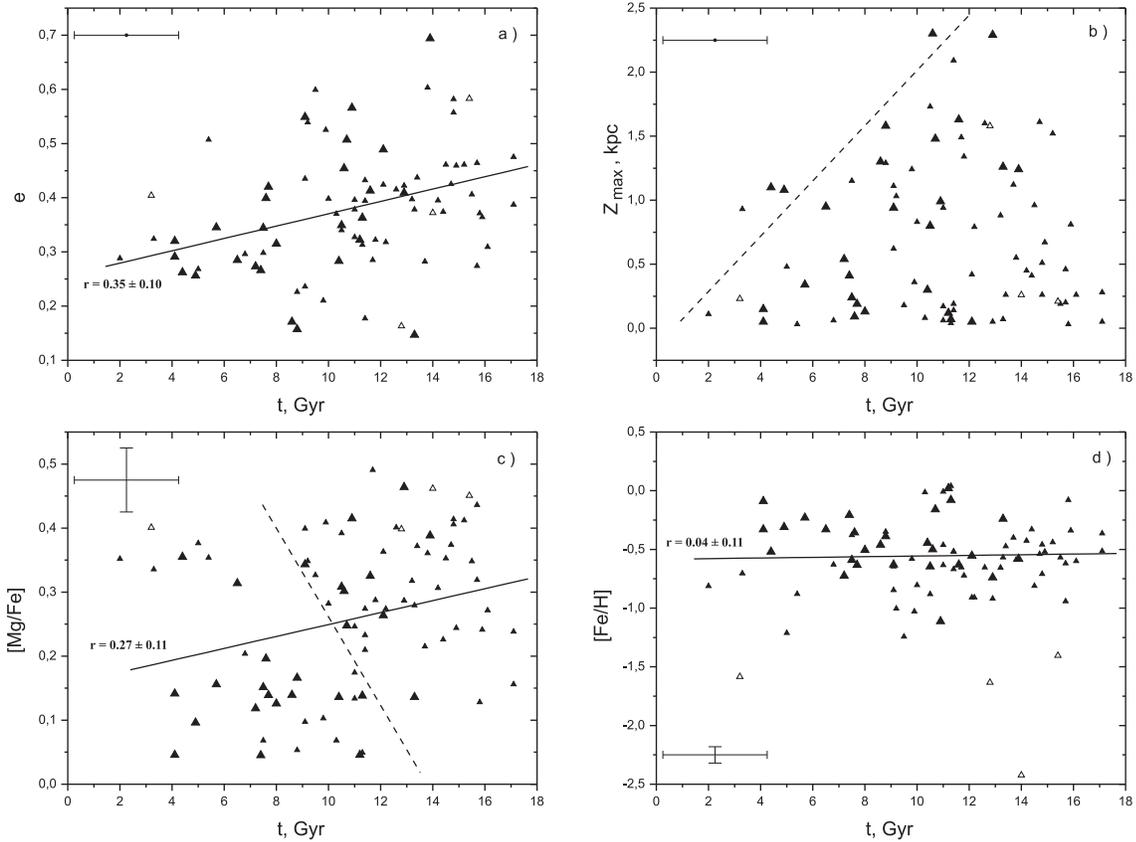}
\caption{Relations of the orbital eccentricities (a), the 
maximum distances of orbital points from the Galactic plane (b), 
and the relative magnesium abundances (c) and metallicities (d) 
in thick-disk stars to the age. The larger triangles denote the 
stars with age errors of less than 2~Gyr. The solid lines represent 
the regression lines. The correlation coefficients are shown. 
See the text for an explanation of the dotted lines in panels (b) 
and (c). The error bars are given.}
\label{fig5}
\end{figure*}

\newpage

\begin{figure*}
\centering
\includegraphics[angle=0,width=0.80\textwidth,clip]{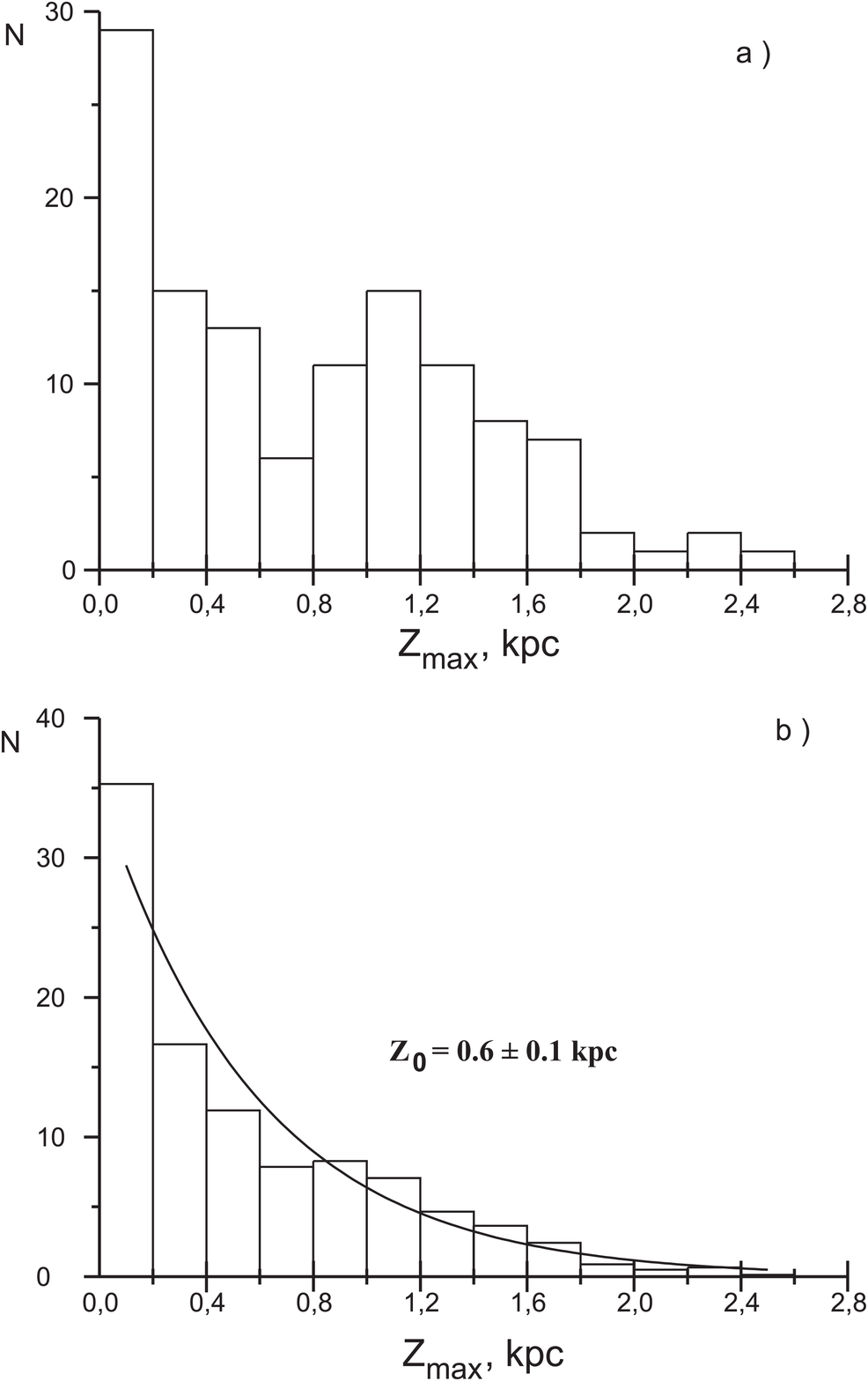}
\caption{Distribution in maximum distance from the Galactic 
plane and (b) reconstructed Z distribution for thick-disk stars.
The solid curve indicates the exponential fit to the distribution, 
and the scale height is indicated together with its uncertainty.}
\label{fig6}
\end{figure*}

\end{document}